\documentclass[twocolumn]{jpsj2} 
\title{Ni-Impurity Effects on Incommensurate Spin Correlations in Superconducting La$_{2-x}$Sr$_{x}$CuO$_{4}$ ($x=0.06$ and $0.07$)
}
\author{Haruhiro~\textsc{Hiraka}
\thanks{E-mail address: hiraka@imr.tohoku.ac.jp}, 
Soichi~\textsc{Ohta}$^{1}$,
Shuich~\textsc{Wakimoto}$^{2}$,
Masaaki~\textsc{Matsuda}$^{2}$
and Kazuyoshi~\textsc{Yamada}
}

\inst{Institute for Materials Research, Tohoku University, Sendai~980-8577\\
$^{1}$Department of Physics, Graduate School of Science, Tohoku University, Sendai~980-8578\\
$^{2}$Quantum Beam Science Directorate, Japan Atomic Energy Agency, Tokai, Ibaraki~319-1195
}

\abst{
Neutron scattering experiments have been carried out to explore Ni-impurity effects on static spin correlations in La$_{2-x}$Sr$_{x}$CuO$_{4}$ (LSCO)  in the vicinity of the superconductor-insulator boundary where both parallel and diagonal spin-density modulations (SDM) coexist at low temperature. Upon dilute Ni substitution the incommensurability decreases for both types of SDM, while the volume fraction of the diagonal (parallel) SDM increases (decreases). 
Subsequent Ni doping induces a bulk three-dimensional antiferromagnetic (AF) order when $x\sim $~Ni concentration.
$T_{\rm N}$ of such the AF order depends on $x$ and seems to disappear at $x\sim 0.1$.
These effects are approximately ascribed by a reduction of mobile holes, and by a transition from the parallel to the diagonal SDM induced by Ni.
}

\kword{La$_{2-x}$Sr$_x$CuO$_{4}$, neutron scattering, spin density modulation, substitution effect, Ni, incommensurate spin correlations}

\begin{document}
\maketitle


\section{Introduction}
\vspace{1mm}
Doped high-$T_{\rm c}$ cuprates,  one of the most fruitful examples of doped Mott insulators,  provide us rich information on the interplay between the doped carriers and the spin correlations commonly underlying on the Cu-O square lattices. Through a series of systematic studies, we have discovered clear relationships between the doping dependence of spin correlation and the onset of the high-$T_{\rm c}$ superconductivity in both underdoped \cite{yamada98} and overdoped superconducting phases.~\cite{wakimoto05}
In the superconducting (SC) phase, the so-called parallel spin-density modulations (P-SDM) commonly exist in hole-doped LSCO. Therefore, the discovery of the so-called diagonal spin-density modulations (D-SDM) in the insulating spin-glass (SG) phase by Wakimoto \textit{et al.} \cite{wakimoto99} strongly suggested a transition of  spin correlation between the D-SDM and P-SDM at the underdoped SG-SC boundary. In fact, the detailed study in the vicinity of the boundary between SG and SC  phases confirmed a D-SDM to P-SDM transition at the boundary.~\cite{wakimoto00,matsuda00,fujita02}

To clarify the origin of the D-SDM we studied the impurity effect in the SG phase.~\cite{matsuda06} The results show that  Ni doping quickly destroys the incommensurability and restores the N\'{e}el ordering, indicating a strong effect on hole localization. This suggests that Ni is doped as Ni$^{3+}$ or as Ni$^{2+}$ with a hole forming a strongly bound state. Therefore, Ni doping reduces the number of mobile or hopping Zhang-Rice (ZR) singlet states around Cu spins by creating localized hole sites near the doped Ni. Then the concentration of the mobile ZR singlet ($x_{\rm eff}$) can be described by the difference between the number of holes and doped Ni ions. In fact, the  $x_{\rm eff}$ dependences of the incommensurability and the onset temperature of the D-SDM for the Ni doped samples can be plot on the same phase diagram without impurities.  This means that the incommensurability in this system is controlled by the number of mobile ZR singlets or mobile holes. 

The similar localization effect around Ni impurities is also observable in lightly doped antiferromagnetic (AF) phase.
Watanabe \textit{et al.} measured electrical resistivity and magnetic susceptibility for dilute hole-doped La$_{2-x}$Sr$_{x}$Cu$_{1-y}$Ni$_{y}$O$_{4}$ (LSCNO) with $x=0.01$.~\cite{watanabe03}  A huge increase of resistivity together with a drastic increase of N\'{e}el temperature ($T_{\rm N}$) was found when doped by a small amount of Ni.
Such the AF order was directly reconfirmed by neutron diffraction using single crystals.~\cite{hiraka05}
In addition, the spin structure was found to switch from La$_2$CuO$_4$-type to La$_2$NiO$_4$-type at $y = 0.05$, suggesting a change from Ni$^{3+}$ ($S=1/2$) to Ni$^{2+}$ ($S=1$).
Machi \textit{et al.} found that such the Ni-enhanced AF order appears for the SG and the underdoped SC phases too from polycrystalline susceptibility.~\cite{machi03}

We further explore the Ni-impurity effects in the SC phase to study whether the strong hole-localization effect by Ni commonly exists in the entire SC phase. In the present neutron scattering experiments we present the results of Ni-impurity effects on the static spin correlations in the SC  La$_{2-x}$Sr$_{x}$CuO$_{4}$ (LSCO) in the vicinity of the SG-SC boundary. Similar to the result in the SG phase we observed a drastic impurity effect in the SC phase.  Upon dilute Ni substitution by $3$\%, both P-SDM and D-SDM considerably shrink in incommensurability ($\delta$), associated with degradation of bulk superconductivity. Subsequent Ni doping induces a bulk three-dimensional AF order with the same spin structure without holes. Based on the hypothesis of the reduction of the effective hole concentration by Ni impurity we propose that the previously studied impurity effects can be simply interpreted.

\section{Experimental}
\vspace{1mm}

\begin{figure}[t]
  \begin{center}
    \leavevmode
    \includegraphics[height=7.5cm]{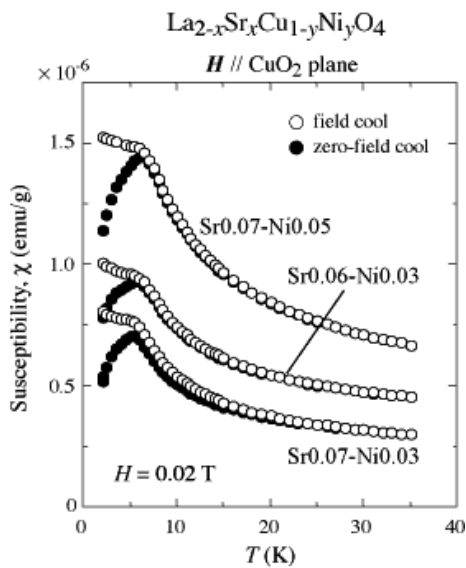}
    \caption{
	Susceptibility data of LSCNO showing Ni-induced spin-glass transitions at $T_{\rm sg}\sim 5$~K.
}
    \label{fig1}
  \end{center}
\end{figure}

\begin{figure}[t]
  \begin{center}
    \leavevmode
    \includegraphics[height=9.5cm]{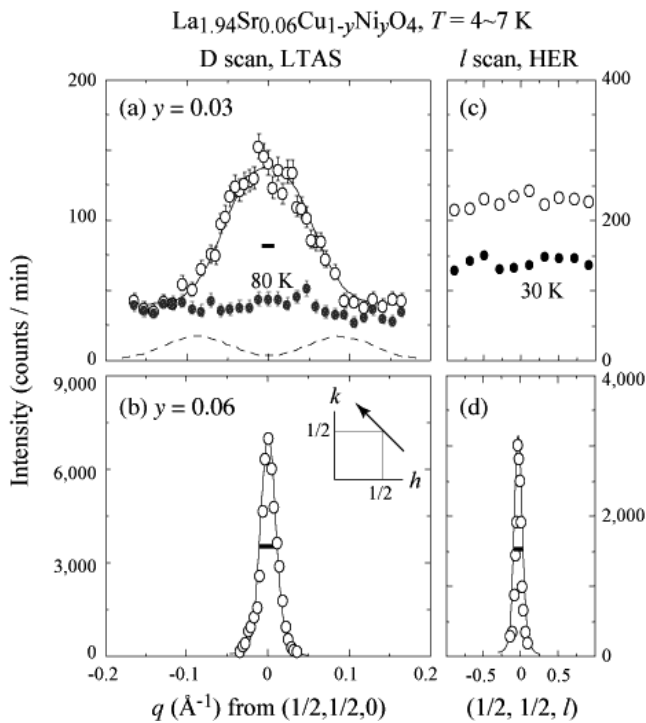}
    \caption{
Magnetic elastic scattering of La$_{1.94}$Sr$_{0.06}$Cu$_{1-y}$Ni$_{y}$O$_{4}$ around $(1/2, 1/2, 0)$ measured (a,b) along the diagonal direction in $(h, k, 0)$ plane and (c,d) along the $l$ direction in $(h, h, l)$ plane, for (a,c) $y=0.03$ and (b,d) $y=0.06$. 
The solid line in (a) is a resolution-convoluted fit to a simple two-peak cross section along the D-scan, while a single Gaussian curved form is assumed in (b) and (d) for the solid lines.
For reference, a resolution-convoluted calculation for $y=0$ without P-SDM is shown by a broken line in (a) using D-SDM parameters of ref.~\cite{fujita02} with an arbitrary intensity scale. 
$Q$ resolutions are shown by short bars. 
}
    \label{fig2}
  \end{center}
\end{figure}

Single crystals of LSCNO of $(x, y) = (0.06, 0.03), (0.06, 0.06), (0.07, 0.03)$ and $(0.07, 0.05)$  were grown by traveling-solvent-floating-zone techniques. The crystals have cylindrical shapes of $4\sim 5$~mm diameter and $\sim 20$~mm length. After our standard heat treatments under oxygen flowing gas, those crystals were characterized chemically by ICP measurements and physically using a SQUID magnetometer. 
In the course of the susceptibility measurements, the volume fraction of bulk susceptibility is found to be strongly suppressed $(< 0.01~\%)$ by Ni, judging from the diamagnetic signals. 
Instead, a spin-glass transition occurs for three crystals $(0.06, 0.03), (0.07, 0.03)$ and $(0.07, 0.05)$ as shown in Fig.~\ref{fig1}, but not for the composition of (0.06, 0.06).

Neutron scattering experiments were carried out on cold-neutron triple-axis spectrometers LTAS and HER installed in the guide hall of JRR-3 at the Japan Atomic Energy Agency (JAEA), and on SPINS in the research reactor of the National Institute for Standard and Technology in U.S.A. Pyrolytic-graphite $(0,0,2)$ reflection was used in both the monochromator and analyzer. Contaminations of higher-order neutrons were sufficiently suppressed by inserting a Be filter into the neutron beam path. Multiple Bragg reflections were removed by tuning incident energies over the range of $4.5\sim 5$~meV. 
Using pseudo-tetragonal lattice parameters of $a^{\ast}\simeq 1.66$~\AA$^{-1} (\approx \sqrt{2} a^{\ast}_{\rm ort}\approx \sqrt{2} b^{\ast}_{\rm ort})$ and $c^{\ast}\simeq 0.48$~\AA$^{-1}$, the scattering process was observed in $(h,k,0)$ and $(h,h,l)$ scattering planes. The horizontal-beam collimation was typically set up to be guide($\sim 20'$)-$80'$-Sample-$80'$-open($\sim 180'$). 
Some parts of sample-quality check determining twin structures were made on AKANE, a thermal-neutron triple-axis spectrometer of Tohoku University installed at JAEA.

All samples studied here consist of twinned crystals, which are naturally caused by the orthorhombic crystal distortion.
The domain distribution was checked by neutron diffraction itself before full measurements of magnetic cross section.
Toward the later section of simulation, we remark here that the single crystals of $(x,y)=(0.06, 0.03)$ and $(0.07, 0.03)$ consist of two types of twinning (or four domains) and one type of twinning (or two domains), respectively.
In these two samples, the domain population is found to be nearly equal because of the comparable peak intensity from each domain.
The orthorhombic lattice parameters are $a^{\ast}_{\rm ort}\simeq 1.18$~\AA$^{-1}$ and $b^{\ast}_{\rm ort}\simeq 1.17$~\AA$^{-1}$ in notation of the low-temperature orthorhombic phase ($Bmab$), and the orthorhombic distortion does not change by current lightly Ni doping [$(b/a)\sim 1.01$].
For simplicity and convenience, we hereafter use a high-temperature tetragonal notation ($I4/mmm$) mainly.

\begin{figure}[t]
  \begin{center}
    \leavevmode
    \includegraphics[height=10cm]{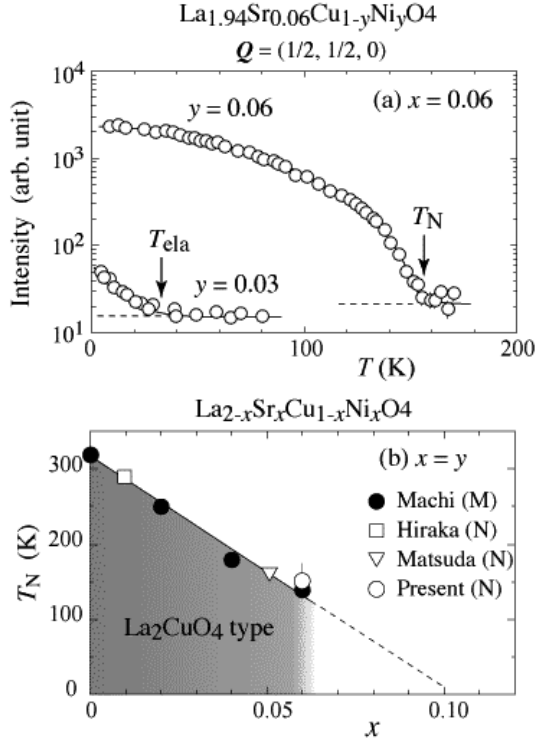}
    \caption{
(a) Thermal evolution of magnetic scattering peak at $(1/2,1/2,0)$ for LSCNO of $(x,y)=(0.06, 0.03)$ and $(0.06, 0.06)$. 
Background levels are shown by broken lines.
(b) Degradation of $T_{\rm N}$ for Ni-induced N\'{e}el ordered state in case of $x=y$, or $x_{\rm eff}(=x-y)=0$. Open and closed symbols stand for data from neutron scattering\cite{hiraka05, matsuda06} and magnetic susceptibility measurements\cite{watanabe03,machi03}, respectively.
}
    \label{fig3}
  \end{center}
\end{figure}

\section{Results}
\vspace{1mm}

Figures~\ref{fig2}(a) and \ref{fig2}(b) show $Q$ spectra around $(1/2,1/2,0)$ along the diagonal-scan (D-scan) direction [the inset of Fig.~\ref{fig2}(b)] for 3\%- and 6\%-Ni doped La$_{1.94}$Sr$_{0.06}$CuO$_{4}$, respectively. For reference, a resolution-convoluted spectrum of LSCO with $x=0.06$ is shown by a broken line in Fig.~\ref{fig2}(a), by using parameters of incommensurate peaks of D-SDM.~\cite{fujita02} The well-defined two-peak structure drastically breaks down upon only 3\%-Ni doping, and a broad commensurate-like peak appears at low temperature [Fig.~\ref{fig2}(a)]. 
Further Ni doping up to $6$\% induces a commensurate sharp peak, which is resolution-limited and much stronger than that of 3\%-Ni doped compound [Fig.~\ref{fig2}(b)]. 
Another difference of magnetic scattering between the two levels of Ni doping appears in $l$ scans [Figs.~\ref{fig2}(c) and \ref{fig2}(d)]. The weak $l$ dependence of the net intensity between $T = 7$~K and $30$~K for $y=0.03$ means a weak interlayer coupling of spins. On the other hand, the sharp resolution-limited peak for $y=0.06$ corresponds to a bulk AF order in the ground state of La$_{1.94}$Sr$_{0.06}$Cu$_{0.94}$Ni$_{0.06}$O$_{4}$.

Figure \ref{fig3}(a) displays temperature dependences of such the Ni-induced AF order and the magnetic diffuse scattering in La$_{1.94}$Sr$_{0.06}$Cu$_{1-y}$Ni$_{y}$O$_{4}$. 
$T_{\rm N} (\simeq 150$~K) of $y=0.06$ well follows the $x$ dependence of $T_{\rm N}$ for La$_{2-x}$Sr$_{x}$Cu$_{1-x}$Ni$_{x}$O$_{4}$~\cite{matsuda06,watanabe03,hiraka05,machi03} in Fig.~\ref{fig3}(b). A linear extrapolation indicates that a bulk AF order can persist up to $x\sim 0.1$. 
As a preliminary step determining the AF spin structure, we measured  three magnetic Bragg reflections $(1,0,0)_{\rm ort}, (0,1,1)_{\rm ort}$ and $(0,1,3)_{\rm ort}$ of $y=0.06$ (not shown). The result shows that the Ni-induced AF spin structure is consistent to that of La$_{2}$CuO$_{4}$~\cite{vaknin87} with a staggered magnetic moment of $0.2\sim 0.3\mu_{\rm B}$/(Cu site) at base temperature. That is, the AF propagation vector is parallel to $[100]_{\rm ort}$, while the spins direct $[010]_{\rm ort}$.

\begin{figure}[t]
  \begin{center}
    \leavevmode
    \includegraphics[height=10cm]{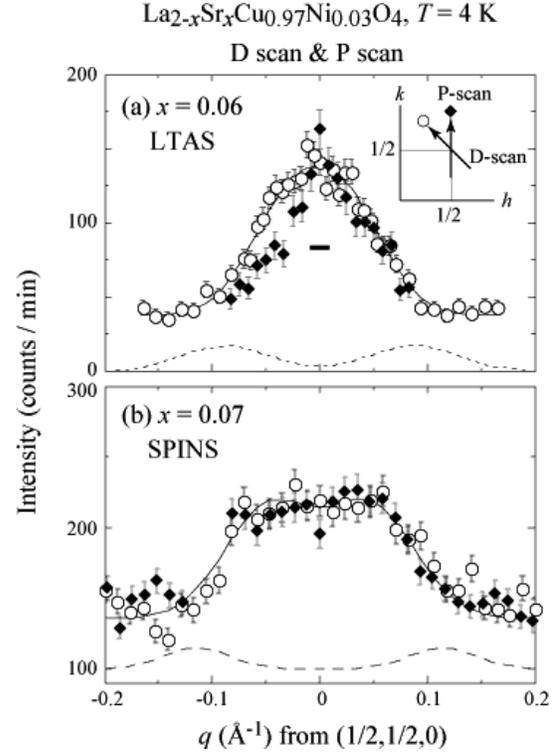}
    \caption{
	Comparison of $\mathbf{Q}$ spectra between the diagonal and parallel scans in La$_{2-x}$Sr$_{x}$Cu$_{0.97}$Ni$_{0.03}$O$_{4}$ for (a) $x=0.06$ on LTAS and (b) $x=0.07$ on SPINS. 
The solid lines are fits to a two-peak structure along the D-scan.
For reference, resolution-convoluted calculations for $y=0$ are shown by broken lines with arbitrary intensity scales, which are using parameters of (a) D-SDM of $x=0.06$~\cite{fujita02} and (b) P-SDM of $x=0.07$~\cite{hiraka01}, respectively.
To see easily, an offset by 100 counts/min is added for the broken line in (b).
The $Q$ resolution is shown by short bar. 
}
    \label{fig4}
  \end{center}
\end{figure}

In the SC phase of LSCO, particularly near the insulator boundary, both D-SDM and P-SDM coexist at low temperature.~\cite{fujita02} In order to clarify Ni-impurity effects on each type of SDM, two types of $\mathbf{Q}$ scan are carried out in the $(h,k,0)$ scattering plane. Figure~\ref{fig4} compares $\mathbf{Q}$ spectra of the D-scan and P-scan for La$_{2-x}$Sr$_{x}$Cu$_{0.97}$Ni$_{0.03}$O$_{4}$ with $x=0.06$ and $0.07$. For $x=0.06$, the P-scan profile is asymmetric about $q=0$, 
possibly due to the domain distribution coupled with a small incommensurability.
By contrast, no clear difference is observable between the D-scan and P-scan for $x=0.07$. 
Besides, the flat-top-like cross section suggests a signature of IC correlations remaining.

\section{Simulation}
\vspace{1mm}

\begin{figure}[t]
 \begin{center}
   \leavevmode
   \includegraphics[height=6cm]{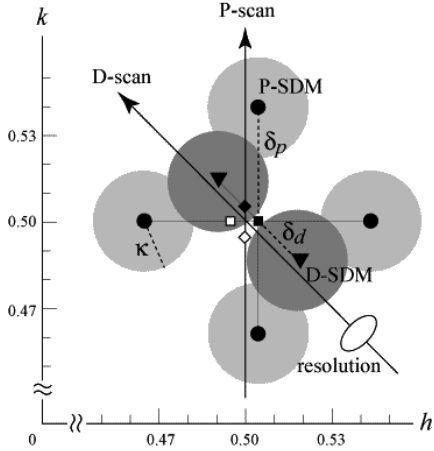}
   \caption{
IC-peak configuration of D-SDM (triangle) and P-SDM (circle), and scan trajectories together with instrumental resolution. 
IC peaks sprouted only from one $(1,0,0)_{\rm ort}$ domain are shown for easy looking.
Parameters used in simulation analysis for $x=0.06$ in Figs.~\ref{fig6}(a)-\ref{fig6}(c) are illustrated in real scale.
Closed [open] squares and diamonds represent the orthorhombic $(1,0,0)_{\rm ort}$ $[(0,1,0)_{\rm ort}]$ position in a four-domain structure caused by two types of twinning. 
}
   \label{fig5}
  \end{center}
\end{figure}

In order to clarify Ni-impurity effects on each type of SDM separately, we focus our discussion on the $\mathbf{Q}$ spectra of La$_{2-x}$Sr$_{x}$Cu$_{0.97}$Ni$_{0.03}$O$_{4}$ in Fig.~\ref{fig4}. Resolution-convoluted simulations have been carried out by taking into the experimentally determined orthorhombic-domain distribution. As schematically shown in Fig.~\ref{fig5} and written in the following cross section, we calculate $\mathbf{Q}$ spectra around $(0.5, 0.5, 0)$ under the existence of both D-SDM and P-SDM with the incommensurability $\delta _{d}$ and $\delta _{p}$, and the peak width $\kappa _{d}$ and $\kappa _{p}$, respectively :
\begin{eqnarray}
I_{\rm sim}(\mathbf{Q}) \sim \sum _{\rm twin}
\biggl[v_{100}\times \bigl( I_{d} + I_{p}\bigr) 
+ v_{010}\times \bigl(\frac{I_{d}}{2} + I_{p} \bigr) \biggr],\label{eq-1}
\\
I_{d}=A_{d}\sum _{\mathbf{Q} _{d}}^{\rm 2~peaks}{\rm exp}\Bigl[-{\rm ln}(2) \Bigl( \bigl|\mathbf{Q}-\mathbf{Q}_{d} \bigr|/\kappa _{d}\Bigr)^{2}\Bigr],\label{eq-2}
\\
I_{p}=A_{p}\sum _{\mathbf{Q}_{p}}^{\rm 4~peaks}
{\rm exp}\Bigl[-{\rm ln}(2) \Bigl( \bigl|\mathbf{Q}-\mathbf{Q}_{p} \bigr|/\kappa _{p}\Bigr)^{2}\Bigr].
\label{eq-3}
\end{eqnarray}
where $v_{100}$ [$v_{010}$] is the volume fraction of orthorhombic $(1,0,0)$ [~$(0,1,0)$~] domain in a twin. $A_{d}$ ($A_{p}$) and $\mathbf{Q_{\it d}}$ ($\mathbf{Q_{\it p}}$) represent the IC peak intensity and the peak position of D-SDM (P-SDM), respectively. The summation with respect to ${\mathbf{Q}_{d}}$ (${\mathbf{Q}_{p}}$) in $I_{d}$ ($I_{p}$) is carried out over two (four) IC peaks, since the D-SDM propagates only along the $[010]_{\rm ort}$ direction.~\cite{wakimoto00} 
Basically, $I_{\rm sim}$ consists of two components. One is a contribution from $(1,0,0)_{\rm ort}$ domain, and the other from $(0,1,0)_{\rm ort}$ domain. For $x=0.06$, because of two types of twin formation in the sample, we repeat this addition for another twin also.
As for the prefactor $(1/2)$ of $I_{d}$ in $(0,1,0)_{\rm ort}$ domain of eq.~(\ref{eq-1}), we referred the experimental result for SG LSCO with $x=0.05$.~\cite{wakimoto00}

Because of the many degrees of freedom in the above model cross section, a proper initial set of SDM parameters is much required to converge our simulation study. To get such the parameters, as a first step, a simple analysis was preparatively done by assuming a two-peak structure along the D-scan direction without P-SDM. A fair agreement is shown in Fig.~\ref{fig4} by curved lines, and the resultant $\delta _{d}$ and $\kappa _{d}$ are listed in Table~\ref{table-1}. Note that the incommensurability definitely decreases with Ni whereas the peak width does not change so much in this preliminary analysis.

\begin{figure}[t]
 \begin{center}
   \leavevmode
   \includegraphics[height=9cm]{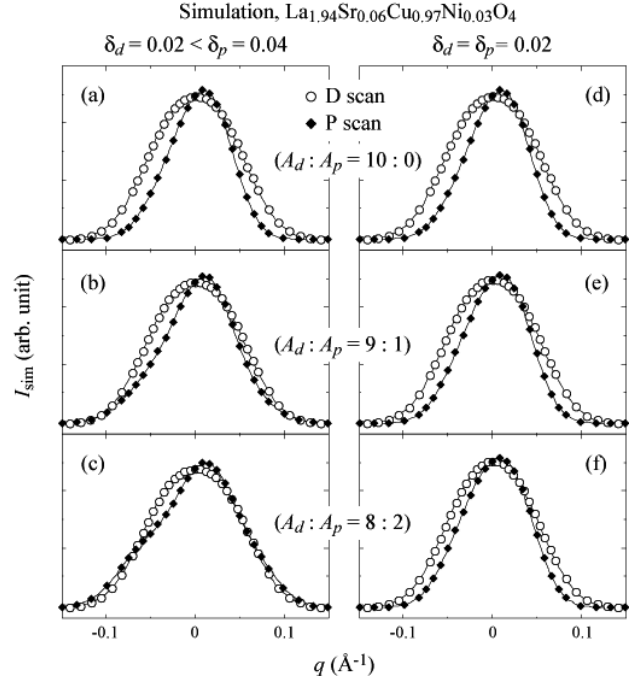}
   \caption{
Two sets of simulation for D-scan and P-scan in La$_{1.94}$Sr$_{0.06}$Cu$_{0.97}$Ni$_{0.03}$O$_{4}$;
(a)-(c) $\delta _{d} < \delta _{p}$ and (d)-(f) $\delta _{d} = \delta _{p}$. 
$A_{d} : A_{p}$ are set to $10 : 0$, $9 : 1$ and $8 : 2$ for (a,d), (b,e) and (c,f), respectively.
}
   \label{fig6}
  \end{center}
\end{figure}

\begin{table}[b]
\caption{Referenced data for coming simulation study, estimated through a simple two-peak fit. For reference, data without Ni are also shown. Note that only the data of $x=0.07$ come from P-SDM.
}
\begin{center}
\vspace{3mm}
\begin{tabular}{cccccl}\hline
$x$ & $y$ & $\delta _{d}$ (r.l.u.) & $\kappa _{d}$ (\AA$^{-1}$) & $\kappa _{d}$ (r.l.u.) \\
\hline
$0.06$ & $0.03$ & $0.017(3)$ & $0.030(4)$ & $0.018(2)$ & present \\
$0.07$ & $0.03$ & $0.029(1)$ & $0.037(4)$ & $0.022(2)$ & present \\
\hline
$0.06$ & $0$ & $0.053(2)$ & $0.039(4)$ & $0.023(2)$ & ref.\cite{fujita02} \\
$\bigl [ 0.07$ & $0$ & $\sim 0.069$ & $\sim 0.037$ & $\sim 0.022$ & ref.\cite{hiraka01}$\bigr]$\\
\hline
\end{tabular}
\end{center}
\label{table-1}
\end{table}%

To simplify the simulation, we fix the peak width to $\kappa _{d}=0.03$~\AA$^{-1}$ for both $x=0.06$ and $0.07$, based on Table~\ref{table-1} and on the data near the SG-SC boundary in LSCO.\cite{fujita02} Besides, we assume $\kappa _{d}=\kappa _{p}$ according to the result of $x=0.06$ without Ni.~\cite{fujita02}
Hence, $I_{\rm sim}$ was calculated for several sets of $\delta _{d}, \delta _{p}, A_{d}$ and $A_{p}$.
The total magnetic intensity of each SDM will be closely related to its magnetic-domain volume.
Thus, we hereafter evaluate the volume-fraction ratio between D-SDM and P-SDM using $V_{d} : V_{p} = 3A_{d} : 8A_{p}$ from eqs.~(\ref{eq-1})-(\ref{eq-3}). 
With an assumption that the spin structure does not change by $3\%$-Ni substitution, we can inspect the magnetic-domain volume before and after Ni doping.
The current results of simulation provide us interesting Ni effect on the spin correlation of this system though it is difficult to precisely determine such the parameters due to the broad feature of the $\mathbf{Q}$ spectra.

\begin{figure}[t]
 \begin{center}
   \leavevmode
   \includegraphics[height=9cm]{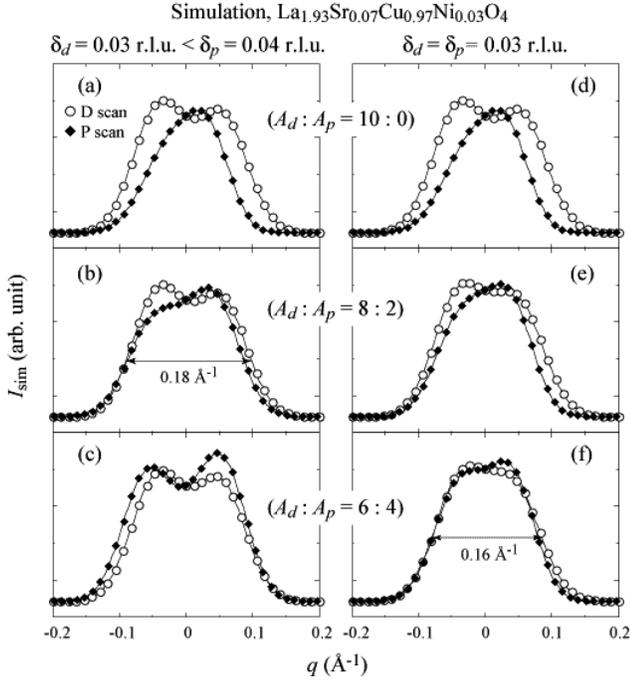}
   \caption{
Two sets of simulation for D-scan and P-scan in La$_{1.93}$Sr$_{0.07}$Cu$_{0.97}$Ni$_{0.03}$O$_{4}$;
(a)-(c) $\delta _{d} < \delta _{p}$ and (d)-(f) $\delta _{d} = \delta _{p}$. 
$A_{d} : A_{p}$ are set to $10 : 0$, $8 : 2$ and $6 : 4$ for (a,d), (b,e) and (c,f), respectively.
FWHM of the flat-top cross section in the D-sacn is shown for (b) and (f).
}
   \label{fig7}
  \end{center}
\end{figure}

Figure~\ref{fig6} shows typical examples of simulation for $x=0.06$. 
In order to reproduce the observed asymmetry in the P-scan [Fig.~\ref{fig4}(a)], 
(1) $\delta _{d}$ should be quite different from $\delta _{p}$ ($\delta _{d}= 0.02$ r.l.u. $<$  $\delta _{p}= 0.04$ r.l.u.) and (2) $A_{d} : A_{p} \sim 9 :1$, as shown in Fig.~\ref{fig6}(b).
This intensity ratio corresponds to $V_{d} : V_{p} \sim 3 : 1$.
These parameters are substantially different from results of Ni-free sample with $x=0.06$;  $\delta _{d}\approx \delta _{p} \simeq 0.05$~r.l.u. and $V_{d} : V_{p} \sim 2: 1$.~\cite{fujita02}

For $x=0.07$ the calculated $\mathbf{Q}$ spectra are less sensitive to the parameters compared to the case of $x=0.06$. Nonetheless  some detailed Ni effect was obtained to explain the specific features seen in Fig.~\ref{fig4}(b); namely, the coincidence of two types of scans and the flat-top-like profiles.
In this sense, Figs.~\ref{fig7}(b) and \ref{fig7}(f) are good candiates to reproduce the experimental data of Fig.~\ref{fig4}(b). 
Further, due to the observed FWHM ($\simeq 0.18$~\AA$^{-1}$) of the flat-top cross section along the D-scan, Fig.~\ref{fig7}(b) looks better than Fig.~\ref{fig7}(f).  
Therefore, (1) $\delta _{d}$ is slightly smaller than $\delta _{p}$ ($\delta _{d}=0.03$ r.l.u. $<$ $\delta _{p}=0.04$ r.l.u.) and (2) $A_{d} : A_{p} \sim 8 :2$.
As a conclusion, $\delta _{p}$ is much smaller than that of Ni-free sample with $x=0.07$ ($\delta _{p}\simeq 0.07$~r.l.u.~\cite{hiraka01}), and the $V_{d}$ ratio ($V_{d} : V_{p} \sim 1.5 : 1$) is quite large, compared to the Ni-free case ($V_{d} : V_{p} \sim 0.7 : 1$).~\cite{comment}

\section{Discussion}
\vspace{1mm}

We studied  Ni-impurity effect on spin correlation in the SC phase near the SG-SC boundary. Similar to the previous results in the AF ordered phase~\cite{hiraka05} and SG phase~\cite{matsuda06}, Ni substitution drastically changes the spin correlation.  
All these facts observed in the wide hole-doping range are commonly explained with an intuitive scenario that Ni and hole couples strongly on the CuO$_2$ planes, thus reducing the number of mobile holes. Indeed, Fig.~\ref{fig7} supports this hypothesis. Surprisingly, the $\delta _{d}$ as well as the $T_{\rm ela}$ in Ni-doped LSCO well follow the data in Ni-free LSCO~\cite{wakimoto99, wakimoto00, matsuda00, fujita02}, when the effective hole concentration is supposed to be $x_{\rm eff}=x - y$. 

\begin{figure}[t]
  \begin{center}
    \leavevmode
    \includegraphics[height=10cm]{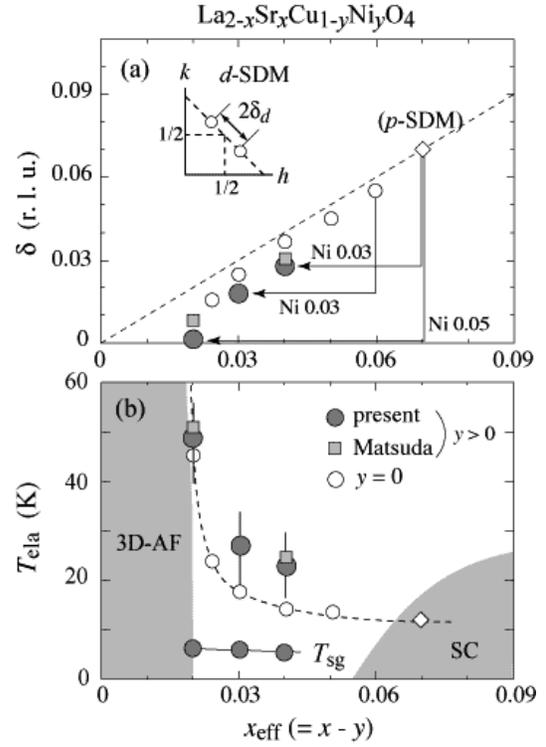}
    \caption{
Plots of (a) incommensurability $\delta _{d}$ and (b) onset temperature $T_{\rm ela}$ of D-SDM peaks against $x_{\rm eff} (= x - y)$ in LSCNO. Filled circles and squares represent data from the SC phase and the SG phase\cite{matsuda06}, respectively. The data of Ni-free LSCO\cite{wakimoto99,wakimoto00,matsuda00,fujita02} are shown by open circles. $T_{\rm sg}$ determined by susceptibility is also plotted in (b).
}
    \label{fig8}
  \end{center}
\end{figure}

In the present case, since the P-SDM and the D-SDM coexist at low temperature, we can further discuss separately the effect of Ni impurity on each SDM phase. The results of analysis taking into the domain distribution provide us interesting information on the spin correlation near the boundary between the SG and SC phases. The previous study without Ni impurities demonstrates a sudden appearance of P-SDM with $\delta _{p}$ = 0.049 (r.l.u in tetragonal unit) upon entering the SC phase with $x > x_{\rm cri} = 0.055$, suggesting a first order transition between the D-SDM and P-SDM.~\cite{fujita02} Since the  maximum value of $T_{\rm c}$ at a given doping $x$ is proportional to the $\delta _{p}$, the first-order transition suggests the existence of a finite minimum value of $T_{\rm c}$. 

Through simulation analysis for the two types of $\mathbf{Q}$ scans in Figs.~\ref{fig6} and \ref{fig7}, specific features of the SDM are newly found. That is, the volume fraction of D-SDM increases by Ni doping, and the incommensurability tends to give a discrepancy between D-SDM and P-SDM. 
Intuitively, the expanding $V_{d}$ is considered as a result of switch from P-SDM to D-SDM at a low temperature by reducing $x_{\rm eff}$ down to bellow $x_{\rm cri}$. The survival of partial P-SDM for $x_{\rm eff}<x_{\rm cri}$ may occur when holes distribute inhomogeneously in CuO$_{2}$ plane and when the hole density exceeds beyond $x_{\rm cri}$ in local. 
Such the microscopic inhomogeneity should be introduced most likely by Ni impurity, and further by supercooling the mobile holes because of crossing boundary of the first-order transition at $x_{\rm cri}$. 
The slightly lower values of $\delta _{d}$ for Ni-doped samples compared to those of Ni-free samples in Fig.~\ref{fig8}(a) as well as the simulation result of $\delta _{d} < \delta _{p}$ supports this scenario.

It is noteworthy that this hole-localization scenario around Ni reasonably explains Ni-impurity effects reported previously in other kinds of physical quantities. We examples three cases below. 
(i) Neutron resonance peak; The well-known magnetic resonant mode appears in inelastic neutron scattering for SC YBa$_{2}$Cu$_{3}$O$_{6+x}$ with resonance energy $E_{\rm r}$.~\cite{rossat91,fong97,dai01} With the help of a relation between $T_{\rm c}$ and the hole concentration ($p$),~\cite{tallon95} $E_{\rm r}$ can be described as a function of $p$. In YBa$_{2}$(Cu$_{0.97}$Ni$_{0.03}$)$_{3}$O$_{7}$, the $E_{\rm r}$ decreases from Ni-free $41$~meV to $35$~meV~\cite{sidis00}. This reduction corresponds to $\Delta p \sim -0.05$ and it agrees semi-quantitatively with the doped Ni-concentration $0.03$. 
(ii) Pseudo-gap; Ni impurity enhances the normal-state pseudo-gap in the $c$-axis optical conductivity of underdoped (Sm,Nd)Ba$_{2}$(Cu$_{1-y}$Ni$_{y}$)$_{3}$O$_{7-\delta}$.~\cite{pimenov05} The increasing pseudo-gap is considered as a natural consequence of the underdoping  by Ni. 
(iii) STM; The SC-coherence peak is little affected by Ni in Bi$_{2}$Sr$_{2}$CaCu$_{2}$O$_{8+\delta}$.~\cite{hudson01} A strong coupling of Ni and hole produces a spin state of $S=1/2$ (either $3d^{7}$ with Ni$^{3+}$, or $3d^{8} \underline{L}$), and then gives a minimum perturbation to the underlying spin-$1/2$ framework. The superconductivity, therefore, will be less damaged by Ni.

Finally, recent our XAFS experiments using synchrotron radiation support our scenario.~\cite{xafs} That is, the valence states of Ni in La$_{1.94}$Sr$_{0.06}$Cu$_{0.97}$Ni$_{0.03}$O$_{4}$ and La$_{1.94}$Sr$_{0.06}$Cu$_{0.94}$Ni$_{0.06}$O$_{4}$ are much different from a Ni$^{2+}$ state, thus indicating most probably either a Ni$^{3+}$ or a strongly hole-bound Ni$^{2+}$ state.


\section*{Acknowledgment}
\vspace{1mm}
We are grateful to J.~Ho for neutron scattering on SPINS at NIST, to S.-H.~Lee, M.~H\"{u}cker, M.~Kofu, M.~Fujita, Y.~Itoh, W.~Koshibae, K.~Tsutsui, T.~Tohoyama and M.~Ogata for stimulating discussions. We also thank K.~Nemoto and N.~Aso for neutron scattering on AKANE and HER at JAERI, and M.~Sakurai for growing the single crystals. The work at Tohoku University was supported by grants from the Ministry of Education, Culture, Sports, Science and Technology. This study was supported by the U.S.-Japan Cooperative Neutron-Scattering Program.  Financial support from the U.S. Department of Energy under Contract DE-AC02-98CH10886 is also gratefully acknowledged.  Work at SPINS is based upon activities supported by the NSF under DMR-9986442. We also acknowledge the U.S. Dept. of Commerce, NIST Center for Neutron Research, for providing the neutron scattering facilities used in this study.




\begin{thebibliography}{99}
\bibitem{comment} This ratio is obtained from elastic scattering measurements using thermal neutron beams (energy resolution$~\sim 2$~meV), and it is qualitatively consistent with the description about $x=0.07$ in ref.~\cite{fujita02}, "... parallel component dominates the diagonal one."
\end{thebibliography}
\end{document}